\documentclass[twocolumn,aps,showpacs,showkeys]{revtex4}
\usepackage{graphicx}              % for \includegraphics

\newcommand{\uck}[1]{\o}

\newcommand{\ket}[1]{\mbox{$|#1\protect\rangle$}}

\def\beq{\begin{equation}}
\def\eeq{\end{equation}}
\def\bea{\begin{eqnarray}}
\def\eea{\end{eqnarray}}

\def\qq{\quad\quad}

\begin{document}

\date{\today}  

\title{Time-Reversed EPR and the Choice of Histories in Quantum Mechanics}

\author{Avshalom C. Elitzur}
\email{avshalom.elitzur@weizmann.ac.il}
\author{Shahar Dolev}
\email{shahar_dolev@email.com}
\affiliation{Unit for Interdisciplinary Studies, Bar-Ilan University, 52900 Ramat-Gan, Israel.}
\author{Anton Zeilinger}
\email{anton.zeilinger@univie.ac.at}
\affiliation{Institut f\"ur Experimentalphysik, Universit\"at Wien, Boltzmanngasse 5, 1090 Wien, Austria.}

\begin{abstract}
When a single photon is split by a beam splitter, its two ``halves'' can entangle two distant atoms into an EPR pair. We discuss a time-reversed analogue of this experiment where two distant sources cooperate so as to emit a single photon. The two ``half photons,'' having interacted with two atoms, can entangle these atoms into an EPR pair once they are detected as a single photon. Entanglement occurs by creating indistinguishabilility between the two mutually exclusive histories of the photon. This indistinguishabilility can be created either at the end of the two histories (by ``erasing" the single photon's path) or at their beginning (by ``erasing" the two atoms' positions). 
\end{abstract}


\maketitle

\section {Introduction}
As peculiar as quantum measurement is known to be, its strangeness is even greater when one tries to determine not merely the state of a system, but its entire {\it history}. Past events are supposed to be unchangeable, and as such the most essential aspect of reality. And yet, when a quantum measurement traces a certain history, it seems to take an active part in the very formation of that history. 

So far, however, this assertion has been merely philosophical. The most notable experiment supporting it, namely, the Einstein-Wheeler ``delayed choice'' experiment (see Sec.~\ref{sec:del_choice}), is equally open to other, less radical interpretations. Could there be a more straightforward experiment, showing that the history observed is retroactively affected by observations carried out much later? In this article we propose a few experiments of this type and discuss their implications. 

\section {The Delayed Choice Experiment} 
\label{sec:del_choice}
\begin{figure}
\centering
\includegraphics[scale=0.7]{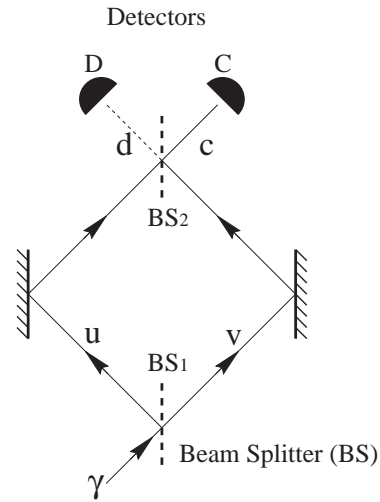}
\caption{Mach-Zehnder Interferometer.}
\label{fig:mzi}
\end{figure}
We shall begin with the ``delayed choice'' experiment. Discussing its limitations will later highlight the advantage of our proposed demonstration of ``choosing history.''

Let a Mach-Zehnder Interferometer (MZI) be large enough such that it takes light a long time to traverse it (Fig.~\ref{fig:mzi}). Due to interference, every single photon traversing this MZI must hit detector $C$. Suppose, however, that, at the last moment, the experimenter decides to pull out $BS_2$. In this case the photon hits either $C$ or $D$ with equal probability. 

What concerned Einstein about this experiment was that the two options given to the experimenter's choice seem to entail two mutually exclusive histories. In the former case the photon seems to have been, {\it all along,} a wave that has traversed both MZI arms and then gave rise to interference. In the latter case the photon must have been -- again, {\it all along} -- a particle: if it has hit $D$ it must have traversed only the right arm, and conversely for $C$. To make the result more impressive, Wheeler \cite{Wheeler78} proposed to perform the experiment on photons coming from outer space, whereby the history thus ``chosen'' is millions-years long. 

However, the delayed choice experiment is not scientific in the full sense of the word, as other explanations are possible within interpretations that do not invoke backward causation. One could, for example, just stick to the observed facts, refrain from any statement about the unobserved past and explain the experiment strictly in terms of wave mechanics or ``collapse.'' 

Can there be an experiment that indicates more strongly that past events are susceptible to the effect of future observation?

\section {Interference between Independent Sources} 
\label{sec:interf_indep_src}
\begin{figure}
\centering
\includegraphics[scale=0.6]{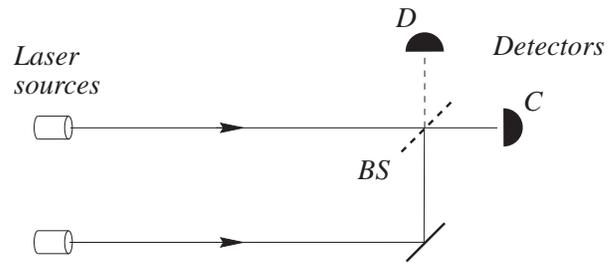}
\caption{A schematic description of Pfleegor-Mandel experiment for interference between two distinct sources.}
\label{fig:mandel}
\end{figure}

Even more striking than the delayed-choice experiment is an effect that was still unknown to Einstein, namely, the interference of light coming from different sources. It was first discovered by Hanbury-Brown and Twiss \cite{HBT57,HBT58}, and later demonstrated at the single-photon level \cite{Pfleegor67,Paul86} (Fig.~\ref{fig:mandel}). It is odd that, although this experiment offends classical notions more than most other experiments known today, it has not yet received appropriate attention. When the radiation involved is of sufficiently low intensity, then even a single particle seems to ``have originated'' from two distant sources. 

We shall first point out two variations of this experiments that highlight its peculiar nature. First, it can have a delayed-choice variant: If the experimenter chooses at the last moment to pull out the BS, a click at detector $C$ will indicate that a single photon has emerged from only one source, namely, the one facing the detector that clicked. If, on the other hand, she leaves the BS in its place, the interference will again indicate that the photon ``has been emitted'' by both sources.

\begin{figure}
\centering
\includegraphics[scale=0.6]{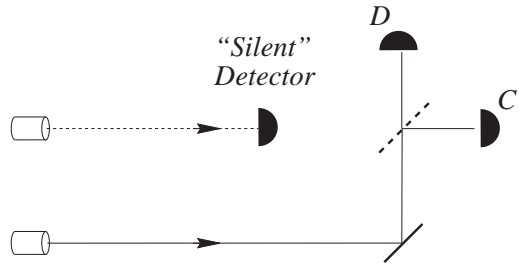}
\caption{A variation of Pfleegor-Mandel experiment, implementing Interaction-Free Measurement.}
\label{fig:mandel-ifm}
\end{figure}

Next consider an Interaction-Free Measurement \cite{EV93} variant of this setting (Fig.~\ref{fig:mandel-ifm}). Assuming that the phase between the sources is fixed for the time of the experiment, it can be arranged that all the photons will reach detector $C$. Now, if an object is placed next to one of the sources, it will occasionally absorb the photon. Therefore, when a photon eventually hits the detector, it is obvious that it has been emitted only from the other, unblocked source. But then, in 50\% of the cases, that photon will emerge from the BS towards the ``dark" detector $D$, thereby indicating that, although it could have originated from only one source, it has somehow sensed the object blocking the other source!

How can two distant sources emit together a single photon? It is instructive to study this effect as a time-reversed version of the familiar case where a single photon is split by a BS and then goes to two distant detectors. In that case, there is an uncertainty as to which detector {\it will} absorb the photon. Similarly, in our case, there is an uncertainty as to which source {\it has} emitted the photon. 

This time-symmetry suggests constructing a new experiment. Consider first the familiar, V-shaped case (one source, two detectors). Such a split photon can entangle two unrelated particles so as to create an EPR pair. For example, two atoms positioned across its two possible paths will become entangled due to the correlation between their ground and excited states. Can the more peculiar, $\Lambda$-shaped case (two sources, one detector) be similarly used to create an inverse EPR? 

\section {Hardy's Hybrid Experiment} 
\label{sec:hardy_exp}
\begin{figure}
\centering
\includegraphics[scale=0.45]{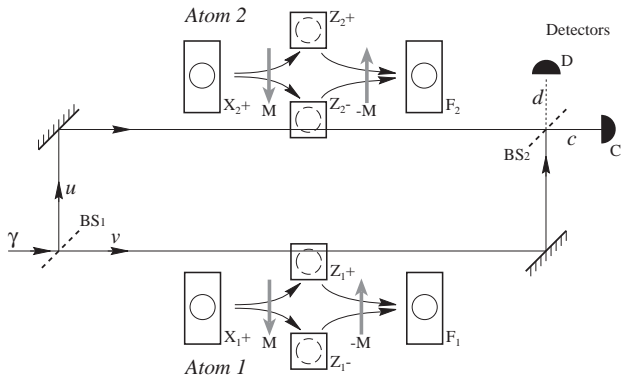}
\caption{Hardy's experiment.}
\label{fig:hardy}
\end{figure}
Before we show how to do that, let us study an experiment due to Hardy \cite{Hardy92a}, in which he has elegantly integrated the peculiarities of the EPR experiment, single-particle interference and the interaction-free measurement -- all in one simple setting. 

Let a single photon traverse a MZI. Let two spin $1 \over 2$ atoms be prepared in the following way: Each atom is first prepared in an up spin-x state ($x^+$) and then split by a non-uniform magnetic field $M$ into its spin-z components. The two components are then carefully put into two boxes $z^+$ and $z^-$ while keeping their superposition state: 
\beq
\Psi=\ket{\gamma} \cdot \frac{1}{\sqrt{2}}(iz^+_1+z^-_1) \cdot \frac{1}{\sqrt{2}}(iz^+_2+z^-_2).
\eeq
The boxes are transparent for the photon but opaque for the atoms. Atom 1's (2's) $z_1^+$ ($z_2^-$) box is positioned across the photon's $v$ ($u$) path in such a way that the photon can pass through the box and interact with the atom inside in a 100\% efficiency. Now let the photon be transmitted by $BS_1$:
\beq
\Psi=\frac{1}{\sqrt{2}^3}(i\ket{u} + \ket{v}) \cdot (iz^+_1+z^-_1)  \cdot (iz^+_2+z^-_2).
\eeq
After the photon was allowed to interact with the atoms, we discard the cases in which absorption occurred (50\%), to get: 
\bea
\Psi&=\frac{1}{\sqrt{2}^3}(&-i\ket{u}z^+_1z^+_2 - \ket{u}z^-_1z^+_2 \\
&&\quad + i\ket{v}z^-_1z^+_2 + \ket{v}z^-_1z^-_2 ). \nonumber
\eea

Now, let photon parts $u$ and $v$ pass through $BS_2$, following the evolution:
\[
\ket{v}\stackrel{BS_2}{\longrightarrow}\frac{1}{\sqrt{2}}\cdot (\ket{d}+i\ket{c}), \qquad
\ket{u}\stackrel{BS_2}{\longrightarrow}\frac{1}{\sqrt{2}}\cdot (\ket{c}+i\ket{d}),
\]
giving:
\bea
\Psi&=\frac{1}{4}(&\ket{d}z^+_1z^+_2 + \ket{d}z^-_1z^-_2 \\
&&+ i\ket{c}z^-_1z^-_2 - i\ket{c}z^+_1z^+_2 -2 \ket{c}z^-_1z^+_2 ). \nonumber
\eea
If we now post-select only the experiments in which the photon was surely disrupted along its way, thereby hitting detector $D$, we get:
\beq
\Psi=\frac{1}{4}\ket{d}(z^+_1z^+_2 + z^-_1z^-_2 ). 
\eeq
Consequently, the atoms, which have never met in the past, become entangled in an EPR-like relation. Unlike the ordinary EPR, where the two particles have interacted earlier, here the only common event in the past is the single photon that has ``visited" both of them. 

In the next section we shall show how to achieve this result even without any common past. Then, the measurement's effect on past evolution will become even more striking.
 
\section {Inverse EPR (``RPE'')} 
\begin{figure}
\centering
\includegraphics[scale=0.55]{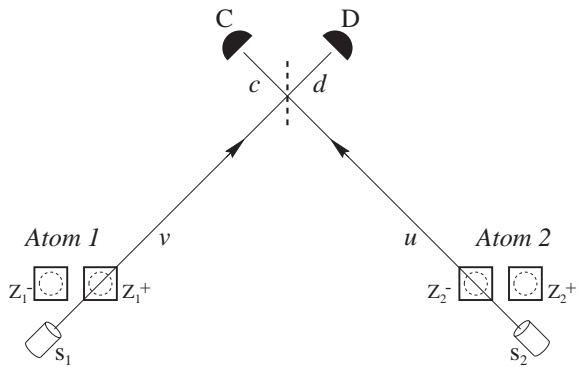}
\caption{Entangling two atoms.}
\label{fig:entangle}
\end{figure}

Let two coherent photon beams be emitted from two distant sources as in Fig.~\ref{fig:entangle}. Let the sources be of sufficiently low intensity such that, on average, one photon is emitted during a given time interval. Let the beams be directed towards an equidistant BS. Two detectors are positioned next to the BS:
\bea
\phi_{\gamma u} &=& p\ket{1}_u+q\ket{0}_u, \\
\phi_{\gamma v} &=& p\ket{1}_v+q\ket{0}_v, \\
\psi_{A 1} &=& {1\over \sqrt 2}(i z^-_1 + z^+_1), \\
\psi_{A 2} &=& {1\over \sqrt 2}(i z^-_2 + z^+_2),
\eea
where $\ket{1}$ denotes a photon state (with probability $p^2$), $\ket{0}$ denotes a state of no photon (with probability $q^2$), $p\ll 1$, and $p^2+q^2=1$.

Since the two sources' radiation is with equal wavelength, a static interference pattern will be manifested by different detection probabilities in each detector. Adjusting the lengths of the photons' paths $v$ and $u$ will modify these probabilities, allowing a state where one detector, $D$, is always silent due to destructive interference, while all the clicks occur at the other detector, $C$, due to constructive interference. 

Notice that each single photon obeys these detection probabilities only if both paths $u$ and $v$, coming from the two distant sources, are open. We shall also presume that the time during which the two sources remain coherent is long enough compared to the experiment's duration, hence we can assume the above phase relation to be fixed.

Next, let two spin-$1 \over 2$ atoms be prepared as in Hardy's experiment (Sec.~\ref{sec:hardy_exp} above) and let each ``half atom'' be placed in one of the possible paths. After the photon was allowed to interact with the atoms, we discard the cases in which absorption occurred (50\%), to get: 
\bea
\Psi&&=\frac{1}{\sqrt{2}^3}(-i\ket{u}z^+_1z^+_2 - \ket{u}z^-_1z^+_2 \\
&&\quad\qq\qq + i\ket{v}z^-_1z^+_2 + \ket{v}z^-_1z^-_2 ). \nonumber
\eea

If we now post-select only the cases in which a single photon reached detector $D$, which means that one of its paths was surely disrupted, we get:
\beq
\Psi=\frac{1}{4}\ket{d}(z^+_1z^+_2 + z^-_1z^-_2 ), 
\eeq
which entangles the two atoms into a full-blown EPR state: 
\[
z^+_1z^+_2 + z^-_1z^-_2.
\]

In other words, tests of Bell's inequality performed on the two atoms will show the same violations observed in the EPR case, indicating that the spin value of each atom depends on the choice of spin direction measured on the other atom, no matter how distant. 

The two photon sources, though unrelated, must still be coherent in order to demonstrate interference. Dropping the coherency requirement would make the EPR inversion even more prominent. This has been accomplished by Cabrillo {\it et. al.} \cite{Cabrillo99} in a different setup, devised for generating pairs of entangled atoms. Their setup involves atoms with three energy levels: two, mutually close ``ground'' states, $\ket{0}$ and $\ket{1}$, and one excited state $\ket{2}$. Two distant such atoms in $\ket{0}$ state are shone by a weak laser beam tuned to the $\ket{0}\rightarrow\ket{2}$ transition energy. If a detector then detects a single photon of the $\ket{2}\rightarrow\ket{1}$ energy, the entangled state $\ket{1}\ket{2}+\ket{2}\ket{1}$ ensues.

Here, in the absence of coherency, one cannot talk about interference. Still, since only one photon is detected, the uncertainty about the photon's origin suffices to make the two atoms entangled, leading eventually to an EPR state. 

Unlike the ordinary EPR generation, where the two particles have interacted earlier, here the only common event lies in the particles' future. These two versions, one involving coherent light and the other with incoherent light, highlight different peculiarities of the inverse EPR, henceforth termed ``RPE.'' We shall discuss their implications below.   

\section {Histories for Choice} 
The ``RPE'' experiment offers several options for studying the way in which measurement determines a history. Consider, first, its delayed-choice aspect, which can be best demonstrated in the incoherent setup of Cabrillio {\it et. al.}: 
\begin{itemize}
\item If the experimenter chooses at the last moment to pull out the BS, then the photon's two possible histories, i.e., ``it originated from the right atom'' and ``it originated from the left atom,'' become distinguishable. Consequently, the photon's ``footprints'' become distinguishable too and no entanglement between the atoms will be observed.

\item Conversely, inserting the BS will entangle the two atoms, even though their interaction with the photon has taken place earlier. In other words, {\it what seems to be the generation of uncertainty only in the observer's mind, gives rise to a testable entanglement in reality.} Unlike the delayed-choice experiment, here the history ``chosen'' leaves observable footprints. 
\end{itemize}

But, in addition to creating uncertainty at the end of the evolution, the coherent version (Fig.~\ref{fig:entangle} ) gives us the freedom to create uncertainty -- or to dissolve it -- also at the beginning of that evolution. For even after the photon was detected at $D$, we can perform two kinds of measurements on the atoms, measurements that will yield conflicting results: 
\begin{itemize}
\item We can measure the position of each atom in one out of the two boxes. In this case, one atom must always be found in the intersecting box, while the other must always reside in the non-intersecting box. Consequently, there is only one possible history for the photon now: {\it It must have taken the path that was not blocked by the atom, never the other, blocked path.} As a result, Bell inequality violations would never be demonstrated by the atoms after this measurement (recall that Bell-inequality statistics cannot be demonstrated on a series of same-spin measurements). Hence, the atoms do not demonstrate non-local correlation. 

\item On the other hand, we can unite the two boxes of each atom using an inverse magnetic field $-M$, and measure the photon's spin along the z axis. Here, we give up the ``which path'' information about the photon. Consequently, Bell-inequality violations would be demonstrated in this case, proving that the photon's two possible histories cooperated so as to entangle the two distant atoms. 
\end{itemize}

All these variants are, in essence, erasure experiments. When we insert the BS in the ``incoherent RPE'' or reunites the atoms in the coherent version, we actually erase the still available information about the photon's two possible histories. Notice, however, that the present erasure experiments (e.g. \cite{Englert91}) demonstrate only the negative result of this information loss, i.e., the disappearance of the interference pattern. The RPE, in contrast, enables erasure to give rise to a positive result, namely, the entanglement of two distant atoms.

{\it ``Nam et ipsa scientia potestas est} (for knowledge itself is power)'' was an old maxim of the ancient Romans, but quantum mechanics rewards one for cases in which {\it ignorance} is generated.

\section {Admit Backward Causation or Abandon Realism?} 
The time-symmetry of quantum theory's formalism is well known \cite{ABL64} and has moreover become the cornerstone of some modern interpretations that render ``affecting the past'' the main characteristic of quantum interaction \cite{ Cramer86,Reznik95}. As early as in 1983, Costa de Beauregard \cite{CdB83} gave a CPT-invariant formulation of the EPR setting that allows a time-reversed EPR. Can we apply such a formulation in our case and assert that the late entangling event, i.e., the detection of the photon, really affects backwards the two histories?

One might argue that our experiment does not really time-reverse the EPR setting because, in order to be sure that Bell's inequality will be violated, the atoms must be measured only after the detection of the entangling photon. Hence, the entangling event still remains in the past of the two correlated atoms. The EPR V shape, so goes the counter-argument, is thus merely flattened rather than turned upside down into a $\Lambda$ shape.  

Notice, however, that the entangling event can lie outside the past light cones of the two atoms' measurements. Here, the argument against backward causation must take the following form: ``The two atoms begin to violate Bell-inequality only at the moment the photon was detected at $D$.'' This statement is relativistically meaningless. By bringing the entangling event itself into spacelike separation with the entangled particles, we actually render both the normal and inverse EPRs equally possible. 

But what does ``affecting the past'' teach us about the nature of time? This question involves a deeper unresolved issue, that of time's apparent ``passage.'' Adherents of the ``Block Universe'' model \cite{Price96}, argue that time's passage is only an illusion. Consequently, all quantum mechanical experiments that seem to involve a last minute decision involve no free choice at all. For example, in the EPR, the experimenter's last-moment decision which spin direction to measure, or, in the ``delayed choice'' experiment, the last-moment decision whether to insert the BS or not, are ``already'' determined in the four-dimensional spacetime. Within this framework, RPE is just as possible as EPR. 

The second alternative is that time has an objective ``flow'' \cite{Prigogine80}. Then, the retroactive entangling effect would occur in some higher time once the ``Now'' has reached the entangling event. 

Both views lie at present outside scientific investigation as both can be neither proved nor disproved. \footnote{However, we have shown elsewhere  that Hawking's information erasure conjecture is more consistent with an objective time ``passage.'' See \cite{Elitzur99} } Hence, a third and a much easier answer to the problem would be dismissing the entire issue by avoiding any reference to objective reality altogether, as in the Copenhagen Interpretation. 

While two of us (AE and SD) tend to the second interpretation and one (AZ) favors the third, we prefer to conclude by pointing out that each side can rely on one of the two giants who have so hotly debated during the first Solvay conferences.

\acknowledgments{We thank Yakir Aharonov, Terry Rudolph and David Tannor for very helpful comments. It is a pleasure to thank the participants of the Workshop on Quantum Measurement Theory and Quantum Information at the Schr\"odinger Institute in Vienna for enlightening discussions.}

%\bibliographystyle{unsrt}

%below, at least one of the bib files will be found!
\bibliography{RPE-final}
%\def\searchbib#1{\IfFileExists{#1.bib}{\bibliography{#1}}{}}
%\searchbib{bib}
%\searchbib{../bib}
%\searchbib{../../bib}
%\searchbib{../../../bib}
\end{document}